\documentclass[10pt,twocolumn,twoside]{IEEEtran}
\usepackage{graphicx}
\usepackage{amsmath}
\usepackage{array}
\newcommand{\PreserveBackslash}[1]{\let\temp=\\#1\let\\=\temp}
\newcolumntype{C}[1]{>{\PreserveBackslash\centering}p{#1}}
\newcolumntype{R}[1]{>{\PreserveBackslash\raggedleft}p{#1}}
\newcolumntype{L}[1]{>{\PreserveBackslash\raggedright}p{#1}}
\usepackage[usenames]{color}
\usepackage{colortbl,booktabs}
\usepackage{tabularx}
\usepackage{multicol}
\usepackage{booktabs}
\usepackage[Symbol]{upgreek}
\usepackage{subfigure}
\usepackage{bm}
\usepackage{url}
\usepackage{amssymb}
\usepackage{cite}
\usepackage{balance}
\usepackage[linesnumbered,boxed,ruled,commentsnumbered]{algorithm2e}
\usepackage{threeparttable}
\usepackage{amsmath}
\usepackage{dcolumn}
\usepackage{multirow}
\usepackage{booktabs}
\usepackage{amsfonts}
\newcolumntype{d}[1]{D{.}{.}{#1}}

\makeatletter
\newcommand*\bigcdot{\mathpalette\bigcdot@{.5}}
\newcommand*\bigcdot@[2]{\mathbin{\vcenter{\hbox{\scalebox{#2}{$\m@th#1\bullet$}}}}}
\makeatother

\begin{document}
	
	\bibliographystyle{IEEEtran} 
	\title{Residual-Aided End-to-End Learning of Communication System without\\ Known Channel}

	\author{Hao Jiang, Shuangkaisheng Bi, Linglong Dai, {\it Fellow, IEEE}, Hao Wang, and Jiankun Zhang 
		
		\thanks{Part of this work has been accepted by IEEE International Conference on Communications (IEEE ICC’21) \cite{icc}.}
		\thanks{H. Jiang, S. Bi, and L. Dai are with the Beijing National Research Center for Information Science and Technology (BNRist) as well as the Department of Electronic Engineering, Tsinghua University, Beijing 100084, China (e-mails: jiang-h18@mails.tsinghua.edu.cn; bsks18@mails.tsinghua.edu.cn; daill@tsinghua.edu.cn).}
		\thanks{H. Wang and J. Zhang are with the Beijing Huawei Technologies Co., Ltd, Beijing 100084, China (e-mails: hunter.wanghao@huawei.com; zhangjiankun4@huawei.com).}
		\thanks{This work was supported in part by the National Key Research and Development Program of China (Grant No. 2020YFB1805005) and in part by the National Natural Science Foundation of China (Grant No. 62031019). {\it (Corresponding author: Linglong Dai.)}}}

	\maketitle
	\begin{abstract}
		 Leveraging powerful deep learning techniques, the end-to-end (E2E) learning of communication system is able to outperform the classical communication system. Unfortunately, this communication system cannot be trained by deep learning without known channel. To deal with this problem, a generative adversarial network (GAN) based training scheme has been recently proposed to imitate the real channel. However, the gradient vanishing and overfitting problems of GAN will result in the serious performance degradation of E2E learning of communication system. To mitigate these two problems, we propose a residual aided GAN (RA-GAN) based training scheme in this paper. Particularly, inspired by the idea of residual learning, we propose a residual generator to mitigate the gradient vanishing problem by realizing a more robust gradient backpropagation. Moreover, to cope with the overfitting problem, we reconstruct the loss function for training by adding a regularizer, which limits the representation ability of RA-GAN. Simulation results show that the trained residual generator has better generation performance than the conventional generator, and the proposed RA-GAN based training scheme can achieve the near-optimal block error rate (BLER) performance with a negligible computational complexity increase in both the theoretical channel model and the ray-tracing based channel dataset. 
		 
	\end{abstract}

    
\section{Introduction}\label{S1}
\IEEEPARstart {T}{hroughout} the history of wireless communications from 1G to 5G, the fundamental wireless system design paradigm remains unchanged, i.e., the whole complicated wireless system can be divided into multiple simpler individual modules, such as source encoder, channel encoder, modulator, demodulator, channel decoder, source decoder, etc. Based on this modular design paradigm, the global optimization of the whole communication system can be approximated by the individual optimization of each module. However, the optimization of each module doesn't mean the global optimization of the whole system~\cite{hoydis1}, e.g., the separate design of modulation and coding is known to be sub-optimal~\cite{e1}. Thus, such a classical design paradigm becomes the bottleneck that limits the globally optimal performance of the wireless communication system.

To break through this bottleneck, the groundbreaking paradigm of end-to-end (E2E) learning of communication system has been recently proposed to jointly optimize the whole system by leveraging powerful deep learning techniques~\cite{hoydis1,hoydis2}. It is well known that deep learning is usually realized by using the multi-layer deep neural network (DNN), in which the adjacent layers are connected by trainable weights. For the E2E learning of communication system, the transmitter and receiver are constructed by fully-connected DNNs, both of which are trained by the standard backpropagation (BP) algorithm to update the trainable weights. In contrast to the classical signal processing algorithms, which are usually complex in wireless communication systems~\cite{hanzo1}, deep learning based E2E learning can realize the modulation and other functions by simple addition and multiplication operations between each layer of the DNN~\cite{dai}. Thus, the E2E learning of communication system could reach or even outperform the conventional system with lower complexity~\cite{hoydis1,zhu1,hoydis3,Bilinear,Capacity,1bit}.


{\color{black}However, the E2E learning of communication system faces a challenging problem, i.e., the transmitter cannot be directly trained by the standard BP algorithm without known channel~\cite{hoydis2}.} To be more specific, in the training process, for the E2E learning of communication system, the transmitter encodes the message by the transmitter DNN. After transmitted through the channel, the received signal is decoded by the receiver DNN. To train the DNNs, the receiver should compute the loss function value, which represents the difference between the receiver output and the transmitted message. After that, the weights of receiver and transmitter DNNs are updated by the BP algorithm, which calculates the gradient of each layer from the derivative of the loss function. The gradient could be obtained directly at the receiver. However, at the transmitter, the gradient is unavailable due to the unknown channel, which blocks the computation of the derivative of the loss function. Consequently, the transmitter could not be trained, which prevents the practical realization of the E2E learning of communication system~\cite{hoydis2}.


\subsection{Prior works}\label{S1a}
To deal with the unknown channel in E2E learning of communication system, different machine learning techniques and architecture design approaches have been recently proposed in the literature ~\cite{hoydis2,raj,hoydis4,wunder,ye1,hoydis5,ye2,gan1,gan2,wgan1}. In the pioneering work~\cite{hoydis2}, a two-phase training solution was proposed. In the first phase, the E2E learning of communication system is trained by assuming a stochastic channel model that is close to the behavior of the practical channel. In the second phase, to compensate for the mismatch of the assumed stochastic channel model and the real channel, only the receiver part is finetuned by supervised learning. Unfortunately, the transmitter is unchangeable in the second phase, which may limit the performance of the system. To improve the system performance, some improved two-phase schemes have been proposed to alternatively train the transmitter and the receiver. These schemes can be generally divided into two categories, i.e., receiver aided schemes and the channel imitation based schemes.

In the first category of receiver aided schemes~\cite{raj,hoydis4,wunder}, the receiver will feedback some information to help the training of the transmitter. Specifically, the simultaneous perturbation stochastic optimization algorithm was used in~\cite{raj} to update the transmitter, and the deep learning technique was used to generate gradient by utilizing the loss function fed back from the receiver. Moreover, reinforcement learning (RL) was utilized at the transmitter in~\cite{hoydis4}, which regarded the loss function value as a reward and the transmitter output as a policy. Then, the transmitter DNN weights could be adjusted according to the reward. However, the policy adaptability was limited by the quantization level and feedback noise. In addition, from the perspective of information theory,~\cite{wunder} applied a neural estimator to estimate the mutual information between the transmitted signal and the received signal, and then optimized the transmitter by maximizing this mutual information. However, this work only considered the simplest case of additive white gaussian noise (AWGN) channel, while the more complex yet practical channel models have not been considered. Note that all schemes mentioned above require a large amount of information transmitted from receiver to transmitter, which increases the system burden.

In the second category of channel imitation based schemes, instead of feeding back a large amount of information from the receiver to the transmitter, some extra modules were added in the system to imitate practical channels. Specifically, a generative adversarial network (GAN) was used in~\cite{ye1} to imitate the real received signal. The GAN contains two parts: the generator and the discriminator, which are both implemented by multi-layer DNNs. In the training process, the generator generated a fake received signal to approximate the distribution of the real received signal, so that the transmitter could be trained reliably through the generator and will not be blocked by the unknown channel. At the same time, the discriminator was used to train the generator to generate the signal as similar to the distribution of the real received signal as possible. In this way, the generator can imitate the real received signal, which builds a bridge for the BP algorithm to calculate the gradient for the transmitter. It was shown that this method can imitate an arbitrary channel and reduce the hardware complexity of transceiver~\cite{ye1,ye2,gan1,gan2,wgan1}.

{\color{black}Unfortunately, there are two problems causing performance degradation for this category of channel imitation based schemes. Firstly, the gradient vanishing problem may happen when training the transmitter through a multi-layer generator.} Secondly, the overfitting problem usually occurs when a mass of parameters are iteratively trained for the transmitter, receiver, generator, and discriminator. These two problems will result in a mismatch between the output of GAN and the real received signal. Consequently, this mismatch will lead to the serious performance degradation of E2E learning of communication system.

\subsection{Our contributions}\label{S1b}
To address the gradient vanishing and overfitting problems of the GAN-based training scheme in E2E learning of communication system, we propose a residual aided GAN (RA-GAN) based training scheme by using the residual neural network (Resnet) to change the layer structure of the generator\footnote{Simulation codes are provided to reproduce the results in this paper:	http://oa.ee.tsinghua.edu.cn/dailinglong/publications/publications.html.
}. The specific contributions of this paper can be summarized as follows.


\begin{itemize}
\item Unlike the conventional generator in GAN to generate the received signal itself, we propose the RA-GAN to generate the \emph{difference} between the transmitted and received signal. Specifically, we build a skip connection that links the input and output layers of the generator to decrease the number of layers from input to output. Since this connection can provide an extra gradient, the proposed RA-GAN is able to mitigate the gradient vanishing problem.

\item {\color{black}We reconstruct the loss function for the proposed RA-GAN to solve the overfitting problem of conventional GAN. Specifically, we introduce the ${\bm l}_2$ regularizer in the loss function to limit the representation ability of the RA-GAN based training scheme for the first time, which is verified to have a better performance than other regularizers.} Note that the increased computational complexity after reconstructing the loss function is negligible compared with simple addition and multiplication operations in DNNs.

\item Simulation results show that the proposed residual generator could generate a much more similar signal to the real received signal than the conventional generator, which verifies the better generation performance of the residual generator. As a result, the RA-GAN based training scheme enables significant block error rate (BLER) performance improvement in both the theoretical channel model and the ray-tracing based channel dataset.

\end{itemize}

\subsection{Organization and notation}\label{S1c}
The remainder of this paper is organized as follows. Section \ref{S2} introduces the preliminaries for E2E learning of communication system. Section \ref{S3} presents the proposed RA-GAN based training scheme. Simulation results are shown in Section \ref{S4}. Finally, the conclusions are summarized in Section \ref{S5}.

{\it Notation}: We denote the column vector by boldface lower-case letters. $\mathcal{CN}\left({0,1}\right)$ is the standard complex Gaussian distribution with mean 0 and variance 1. ${\mathbb{R}}_+^\emph{M}$ denotes $M$ dimensional positive real number. $\mathbb{E}\left\lbrace\cdot\right\rbrace$ denotes the expectation. ${\bf I}_n$ denotes the identity matrix of size $n$. $\left|\cdot\right|$ and $\left\|\cdot\right\|^2$ denote the number of weight parameter and $\bm l_2$ regularization, respectively. For functions  ${\bm{f_x}\in\mathbb{R}}^n$ and  ${\bm{f_y}\in\mathbb{R}}^k$ with variable ${\bm{x},\bm{y}\in\mathbb{R}}^m$, $\frac{\partial \bm{f_y}}{\partial \bm{f_x}}\in\mathbb{R}^{k\times n}$, $\frac{\partial \bm{y}}{\partial \bm{x}}\in\mathbb{R}^{m\times m}$, and $\nabla_{\bm{x}}\bm{f_x}\in\mathbb{R}^{n\times m}$ are their gradient matrices.

\section{Preliminaries for E2E Learning of Communication System}\label{S2}

\begin{figure}[tbp]
	\begin{center}
		\subfigure[The architecture of the classical wireless communication system, which includes source encoder, channel encoder, modulator, channel, demodulator, channel decoder, and source decoder.]{
		\hspace*{0mm}\includegraphics[width=1\linewidth]{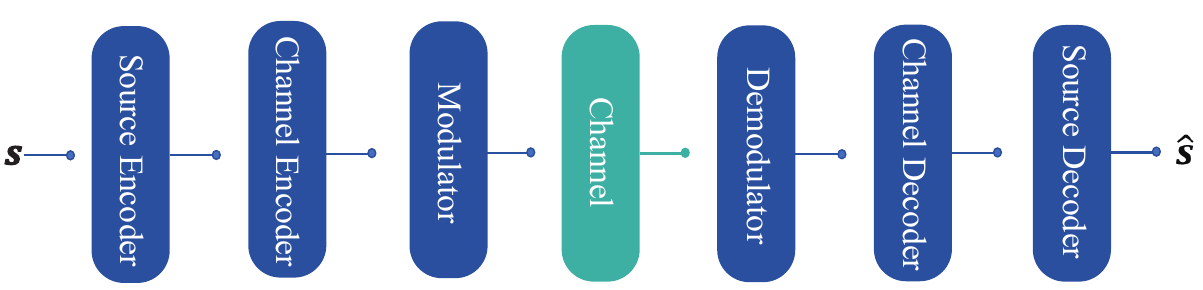}
		}%
	
	    \subfigure[The architecture of E2E learning of communication system~\cite{hoydis1}.]{
	    \hspace*{0mm}\includegraphics[width=1\linewidth]{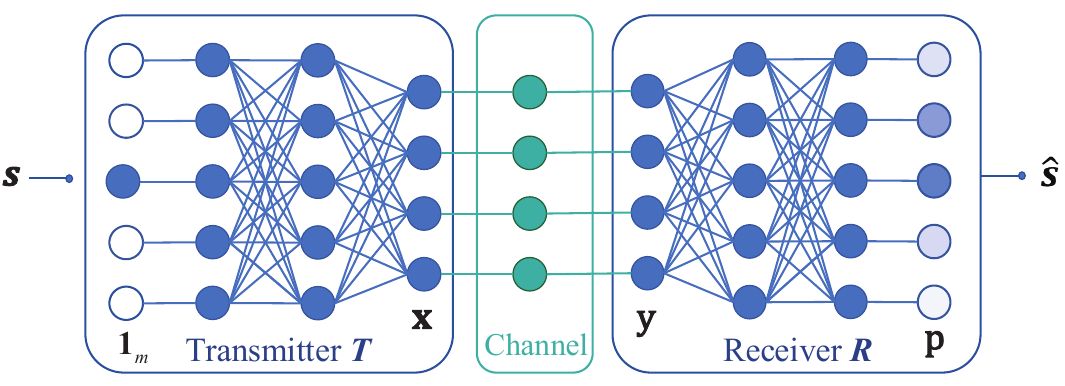}
	    }%
	\end{center}
	\vspace*{-2mm}\caption{Architecture comparison between the classical communication system and end-to-end learning of communication system.} \label{FIG.1}
	\vspace*{+2mm}
\end{figure}
In this section, we will first introduce the principles of E2E learning of communication system and the corresponding problem caused by unknown channel. Then, we will show how GAN could solve this problem, where the associated problems of gradient vanishing and overfitting will be discussed.

\subsection{The principle of end-to-end communication system}
The architecture comparison between the classical communication system and E2E learning of communication system is shown in Fig. \ref{FIG.1}. In the classical wireless communication system, the whole complicated system is divided into multiple individual function modules such as source encoder, channel encoder, modulator, channel, demodulator, channel decoder, source decoder, etc. In contrast, the E2E learning of communication system is only composed of three parts: transmitter, channel, and receiver. Both transmitter $\bm T$ and receiver $\bm R$ are implemented by multi-layer DNNs, with the trainable weights denoted by $\bm{\theta_T}$ and $\bm{\theta_R}$, respectively. Note that the input information $\bm s$ to the transmitter is mapped to a one-hot vector ${\bf{1}}_m$, which is an ${M}$-dimensional vector taken from set $\mathcal{M}$, where only the $m$-th element is one, while the rest ${M-1}$ elements are zeros. {\color{black}Then, the transmitter acts as a function ${\bm{f}_{\bm{\theta_T}}: \mathcal{M} \mapsto \mathbb{C}}^n$, which maps the one-hot vector ${\bf{1}}_m$ to the signal ${\bf{x}\in\mathbb{C}}^n$ to be transmitted through $n$ time slots.} Correspondingly, the receiver acts as a function ${\bm{f}_{\bm{\theta_R}}:\mathbb{C}}^n\mapsto \{\bf{p}\in{\mathbb{R}}_+^\emph{M}\ |\ \sum_{\mathit{i=}\mathrm{1}}^\emph{M}\mathit{p_i}=\mathrm{1}\}$, which maps the received signal ${\bf{y}\in\mathbb{C}}^n$ to a probability vector $\bf{p}\in{\mathbb{R}}_+^\emph{M}$. The final decision of $\bm{\hat s}$ will correspond to the one-hot vector ${\bf{1}}_{\widehat m}$, where ${\widehat m}$ is the index of the maximal element in the probability vector $\bf p$. {\color{black}The block error rate is defined as $P_e=\frac{1}{M}\sum_{\bm s}{\rm Pr}({{\bm{\hat s}}\neq {\bm s}}|{\bm s})$~\cite{hoydis2}, which denotes the average error rate {\color{black} when} transmitting the different message $\bm s$.} In general, the the transmitter hardware introduces the power constraint on the transmitted signal $\bf{x}$, i.e., $\left\|\bf{x}\right\|^2=1$. {\color{black}The purpose of {\color{black}the} transmitter-receiver is to recover the message ${\bf{1}}_m$ as accurately as possible from the received signal ${\bf y}=\bf h\bigcdot\bf{x+w}$, where ${\bf h}\in{\mathbb{C}^n}$ is assumed as the block fading channel. {\color{black}The channel coefficients in block fading channel change independently from one time slot to another. The ${\bf w}\in{\mathbb{C}}^n$ is Gaussian noise.} In detail, the received signal ${\bf y}=[y_1,y_2,\cdot\cdot\cdot,y_n]^T$ at each time slot could be calculated by $y_i=h_ix_i+w_i,i=1,2,\cdot\cdot\cdot,n$. Without loss of generality, we consider the slow fading channel, where the channel keep unchanged in $n$ time slots, i.e., $h=h_i,i=1,2,\cdot\cdot\cdot,n$. Correspondingly, we simplify the representation of the received signal by ${\bf y} = h{\bf x}+{\bf w}$. Moreover, the channel could be denoted by conditional probability $p_{h}({\bf y|x})$.}

In order to get the optimal weights $\bm{\theta_T^*}$ and $\bm{\theta_R^*}$ for transmitter and receiver, we should train the transmitter DNN and receiver DNN. In the training process, the transmitted information is known in receiver, which could be generated by using the same random seed in transmitter and receiver. Then, the difference between the transmitted one-hot vector ${\bf{1}}_m$ and the recovered probability vector ${\bf{p}}$ is measured by a loss function~\cite{hoydis4} as follows:
\begin{align}\label{eq1}
{{\mathcal{L}}\bm{\left(\theta_T,\theta_R},{\mathcal{H}}\bm{\right)}}&\triangleq\mathbb{E}_{\mathcal{H}}\left\lbrace \int l{\left(\bm{f}_{\bm{\theta_R}}\left({\bf{y}}\right),{\bf{1}}_m\right)}{p_{h\in{\mathcal{H}}}\left({\bf{y}}|\bm{\mathit{f}}_{\bm{\theta_T}}\left({\bf{1}}_m\right)\right)} d\bf{y}\right\rbrace \notag \\
&\approx\frac{1}{B}\sum_{\mathit{i=}\mathrm{1}}^\emph{B}l\left({\bm{f}}_{\bm{\theta_R}}\left({\bf{y}}^{\left(\mathit{i}\right)}\right),{\bf{1}}_m^{\left(\mathit{i}\right)} \right)\notag \\
&=\frac{1}{B}\sum_{\mathit{i=}\mathrm{1}}^\emph{B}l\left({\bf{p}}^{\left(\mathit{i}\right)},{\bf{1}}_m^{\left(\mathit{i}\right)} \right),
\end{align}
where ${{\mathcal{H}}=\{{h}^{(1)},\cdot\cdot\cdot,{h}^{(B)}\}}$ is the training set of the channel, 
\begin{equation}\label{eq11} \!l\left({\bf{p}},{\bf{1}}_m\right)=-\sum_{\mathit{j=}\mathrm{1}}^{M}\left({\bf{1}}_m\right)_{j}\ln{{\bf{p}}_j}\!+\!\left(1-\left({\bf{1}}_m\right)_{j}\right)\ln\left(1-{\bf{p}}_j\right) 
\end{equation}
is the cross-entropy (CE) loss function representing the distance between one-hot vector $\bf{1_m}$ and probability vector $\bf{p}$, ${B}$ is the batch size (the number of training samples to estimate the loss function). The ${\bf{p}}^{\left(\mathit{i}\right)}$, ${\bf{y}}^{\left(\mathit{i}\right)}$, and $\bf{1}_m^{\left(\mathit{i}\right)}$ are the $i$-th probability vector, received signal, and training sample, respectively. Next, to  update the weights $\bm{\theta_T}$ and $\bm{\theta_R}$ for transmitter and receiver DNNs,  the gradient of the loss function ${{\mathcal{L}}\bm{\left(\theta_T,\theta_R,}{\mathcal{H}}\bm{\right)}}$ in (\ref{eq1}) is required to be calculated by the classical BP algorithm.
 However, from (\ref{eq1}), only $\bm{\theta_R}$ could be updated by applying the gradient defined as follows:
\begin{equation}\label{eq2}
\nabla_{\bm{\theta_R}}\widetilde{\mathcal{L}}\left({\bm{\theta_R}}\right) = \frac{1}{B}\sum_{\mathit{i=}\mathrm{1}}^\emph{B}\nabla_{\bm{\theta_R}}l\left(\bm{f}_{\bm{\theta_R}}\left({\bf{y}}^{\left(\mathit{i}\right)}\right),{\bf{1}}_m^{\left(\mathit{i}\right)} \right),
\end{equation}
where $\widetilde{\mathcal{L}}$ is an approximation of the loss function, which could be computed from (\ref{eq1}). To fully exploit the performance of E2E learning, the transmitter DNN weights $\bm{\theta_T}$ also need to be optimized~\cite{hoydis1}. However, the gradient $\nabla_{\bm{\theta_T}}\widetilde{\mathcal{L}}$ with respect to $\bm{\theta_T}$ is unavailable~\cite{hoydis2}, since the unknown channel ${h}$ blocks the backpropagation procedure as follows:
\begin{align}\label{eq3}
\nabla_{\bm{\theta_T}}\widetilde{\mathcal{L}}\left({\bm{\theta_T}}\right) 
&=\frac{1}{B}\sum_{\mathit{i=}\mathrm{1}}^\emph{B}\nabla_{\bm{\theta_T}}l\left(\bm{f}_{\bm{\theta_T}}\left({\bf{y}}^{\left(\mathit{i}\right)}\right),{\bf{1}}_m^{\left(\mathit{i}\right)} \right)\notag \\
&=\frac{1}{B}\sum_{\mathit{i=}\mathrm{1}}^\emph{B}\frac{\partial l}{\partial \bm{f}_{\bm{\theta_R}}}\frac{\partial \bm{f}_{\bm{\theta_R}}}{\partial {\bf{y}}^{\left(\mathit{i}\right)}}\frac{\partial {\bf{y}}^{\left(\mathit{i}\right)}}{\partial {\bf{x}}^{\left(\mathit{i}\right)}}\nabla_{\bm{\theta_T}}\bm{f}_{\bm{\theta_T}}\left({\bf{1}}_m^{\left(\mathit{i}\right)}\right)\notag \\
&=\frac{1}{B}\sum_{\mathit{i=}\mathrm{1}}^\emph{B}h^{(i)}\frac{\partial l}{\partial \bm{f}_{\bm{\theta_R}}}\frac{\partial \bm{f}_{\bm{\theta_R}}}{\partial {\bf{y}}^{\left(\mathit{i}\right)}}\bf{I}_n\nabla_{\bm{\theta_T}}\bm{f}_{\bm{\theta_T}}\left({\bf{1}}_m^{\left(\mathit{i}\right)}\right).
\end{align}

To address this problem, a GAN based training scheme was proposed in~\cite{ye1} to generate a surrogate gradient, which will be discussed in the next subsection.
\begin{figure}[tp]
	\begin{center}
		\hspace*{0mm}\includegraphics[width=1\linewidth]{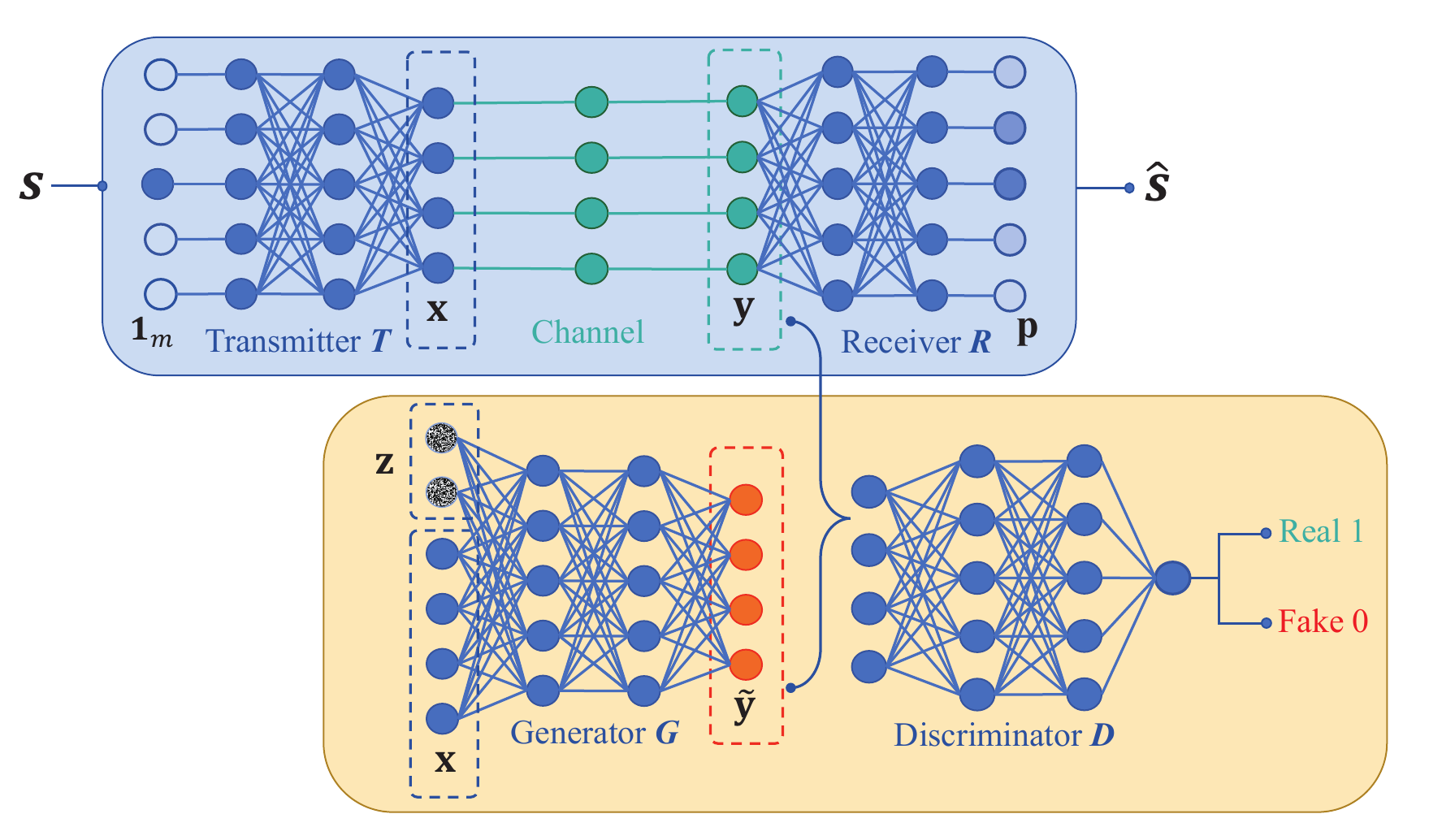}
	\end{center}
	\vspace*{-4mm}\caption{{GAN based training schme~\cite{ye1}.}} \label{FIG.2}
	\vspace*{+1mm}
\end{figure} 
\subsection{GAN based training scheme}\label{2b}

In order to update the transmitter DNN weights $\bm{\theta_T}$, a GAN was used to produce the surrogate gradient~\cite{ye1} as shown in Fig. \ref{FIG.2}, which could generate the signal as similar to the real received signal as possible. Generally, a GAN contains a generator $\bm G$ and a discriminator $\bm D$, both of which are implemented by multi-layer DNNs, with trainable weights denoted by $\bm{\theta_G}$ and $\bm{\theta_D}$, respectively. {\color{black}The generator ${\bm{f}_{\bm{\theta_G}}}: {\mathbb{C}}^{2n} \mapsto {\mathbb{C}}^n$ produces fake received signal ${\bf{\widetilde{y}}}\in{\mathbb{C}}^{n}$ according to the transmitted signal ${\bf{x}}\in{\mathbb{C}}^{n}$ and random noise ${\bf{z}}\in{\mathbb{C}}^{n}$ following the standard Gaussian distribution. In order to imitate the randomness of the channel, the random input $\bf z$ sampling from a Gaussian distribution is required, which make the generator $\bm G$ produce different output after giving $\bf x$. Accordingly, the generator could produce a distribution that approximates the real received signal distribution.} At the same time, the discriminator ${\bm{f}_{\bm{\theta_D}}}: {\mathbb{C}}^n \mapsto \left(0,1\right)$ is used to train the generator to generate the signal as similar to the distribution of the real received signal as possible.
 
 The objective of the discriminator $\bm{D}$ is to accurately distinguish real and fake received signals. Particularly, if the input data of the discriminator is the real received signal ${\bf{y}}$, the expected output of discriminator is 1. On the contrary, if the input data is the fake received signal ${\bf\widetilde{y}}$ generated by the generator, the expected output is 0. For the generator $\bm{G}$, in order to generate a signal as similar to the real received signal as possible, its output $\bf\widetilde{y}$ must make the discriminator output  $\bm{f}_{\bm{\theta_D}}\left({\bf\widetilde{y}}\right)$ as close to 1 as possible. Based on the working procedure of GAN discussed above, the generator weights $\bm{\theta_D}$ and the discriminator weights $\bm{\theta_G}$ are alternately updated according to the following two loss functions: 
 \begin{equation}\label{eq4}
 \widetilde{\mathcal{L}}\left(\bm{\theta_D}\right) = \frac{1}{B}\sum_{\mathit{i=}\mathrm{1}}^\emph{B}\left\lbrace l\left(\bm{f}_{\bm{\theta_D}}\left({\bf{y}}^{\left(\mathit{i}\right)}\right),1 \right)+l\left(\bm{f}_{\bm{\theta_D}}\left({\bf\widetilde{y}}^{\left(\mathit{i}\right)}\right),0 \right)\right\rbrace,
 \end{equation}
 \begin{equation}\label{eq5}
\widetilde{\mathcal{L}}\left(\bm{\theta_G}\right)= \frac{1}{B}\sum_{\mathit{i=}\mathrm{1}}^{B}l\left(\bm{f}_{\bm{\theta_D}}\left({\bm{f}_{\bm{\theta_G}}\left({\bf{x}}^{\left(\mathit{i}\right)},{\bf{z}}^{\left(\mathit{i}\right)}\right)}\right),1\right),
 \end{equation}
where the function $l(\cdot)$ is defined similarly to (\ref{eq11}). The discriminator loss function (\ref{eq4}) contains two items. Specifically, the first item in braces of (\ref{eq4}) denotes the loss function of the real received input $\bf{y}$, while the second item denotes the loss function of the fake received input $\bf\widetilde{y}$. Then, the gradients could be computed by $\nabla_{\bm{\theta_G}}\widetilde{\mathcal{L}}\left({\bm{\theta_G}}\right)$ and $\nabla_{\bm{\theta_D}}\widetilde{\mathcal{L}}\left({\bm{\theta_D}}\right)$, and Adam gradient descent algorithm~\cite{adam} can be used to minimize the loss functions (\ref{eq4}) and (\ref{eq5}). The training process will stop when GAN reaches the Nash equilibrium, i.e., when the discriminator output is nearly 0.5, which means the real and fake received signals can not be distinguished anymore. Since the generator can be trained to imitate the real received signal, the surrogate gradient as close to the expected gradient (\ref{eq3}) as possible could be passed back through the link of transmitter-generator-receiver as follows:
 \begin{align}\label{eq6}
\nabla_{\bm{\theta_T}}\widetilde{\mathcal{L}}\left(\bm{\theta_T}\right)&= \frac{1}{B}\sum_{\mathit{i=}\mathrm{1}}^{B}\nabla_{\bm{\theta_T}}l\left(\bm{f}_{\bm{\theta_R}}\left({\bm{f}_{\bm{\theta_G}}\left(\bm{f}_{\bm{\theta_T}}\left({{\bf{1}}_m^{\left(\mathit{i}\right)}}\right),{\bf{z}}^{\left(\mathit{i}\right)}\right)}\right),{\bf{1}}_m^{\left(\mathit{i}\right)}\right) \notag \\
&=\frac{1}{B}\sum_{\mathit{i=}\mathrm{1}}^{B}\frac{\partial l}{\partial \bm{f}_{\bm{\theta_R}}}\frac{\partial \bm{f}_{\bm{\theta_R}}}{\partial \bm{f}_{\bm{\theta_G}}}\frac{\partial \bm{f}_{\bm{\theta_G}}}{\partial \bm{f}_{\bm{\theta_T}}}\nabla_{\bm{\theta_T}}\bm{f}_{\bm{\theta_T}}\left({\bf{1}}_m^{\left(\mathit{i}\right)}\right)\\
&{\color{black}=\frac{1}{B}\sum_{\mathit{i=}\mathrm{1}}^{B}\frac{\partial l}{\partial \bf{p}^{(\mathit i)}}\frac{\partial \bf{p}^{(\mathit i)}}{\partial \bf{\widetilde y}^{(\mathit i)}}\frac{\partial \bf{\widetilde y}^{(\mathit i)}}{\partial \bf{x}^{(\mathit i)}}\nabla_{\bm{\theta_T}}\bm{f}_{\bm{\theta_T}}\left({\bf{1}}_m^{\left(\mathit{i}\right)}\right)}, \notag
\end{align}
{\color{black}where the $\bm{f}_{\bm{\theta_R}}, \bm{f}_{\bm{\theta_G}}$, and $\bm{f}_{\bm{\theta_T}}$ are used to {\color{black}denote} the output of the receiver, generator, and transmitter, respectively.} Due to the different objectives for transmitter, receiver, generator, and discriminator, the modules are iteratively trained, i.e., when we train one module, the weights of other
modules remain unchanged.

However, it is well known that the training instability problem that limits the performance of GAN~\cite{Wasserstein}. Specifically, in the GAN based training scheme of the E2E learning of communication system, the gradient vanishing problem happens in a multi-layer generator, which makes the transmitter very difficult to be trained. Moreover, the overfitting problem always occurs because a mass of weights are iteratively trained for the transmitter, receiver, generator, and discriminator, which makes the system weights easily overfit to the batch training data. These two problems will result in a serious mismatch between the output of GAN and the real received signal. Consequently, this mismatch will result in the serious performance degradation of E2E learning of communication system. {\color{black}To address the gradient vanishing and overfitting problems of GAN based training scheme in E2E learning of communication system, we will propose a residual aided GAN (RA-GAN) based training scheme to train the transmitter indirectly in the next section.}


\section{RA-GAN Based Training Scheme}\label{S3}
The training of the transmitter is a challenging task because of the unknown channel. According to the GAN based training scheme, a surrogate gradient is produced in (\ref{eq6}) to update the transmitter. However, the generator output distribution  ${p_{\tilde{h}}(\bf\widetilde{y}|\bf{x})}$ is inconsistent with the real received signal distribution ${p_h(\bf{y}|\bf{x})}$ due to the gradient vanishing and overfitting problems. Therefore, we propose an RA-GAN based training scheme in this section to address these two problems.


\subsection{Residual learning to mitigate gradient vanishing}\label{S3a}
\begin{figure}[tp]
	\begin{center}
		\hspace*{0mm}\includegraphics[width=1\linewidth]{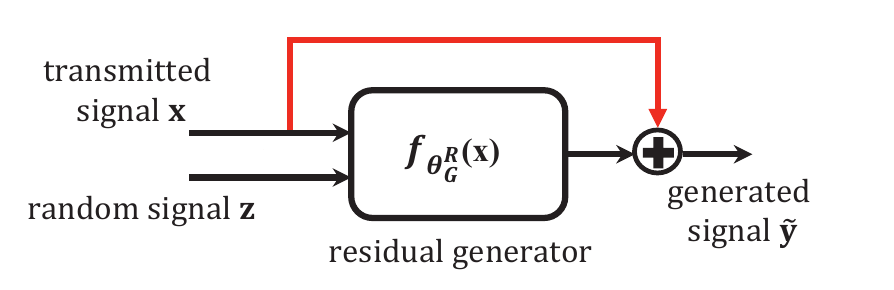}
	\end{center}
	\vspace*{-2mm}\caption{{\color{black}The residual generator in proposed RA-GAN.}} \label{FIG.3}
	\vspace*{2mm}
\end{figure}

{\color{black}In conventional GAN, the multi-layer generator always feeds forward the variables layer-by-layer and feeds back the gradients layer-by-layer. However, as the generator depth increases, the gradient may become very small. This is caused by the fact that the gradient is obtained by multiplying the partial derivatives of loss functions layer-by-layer in the classical BP algorithm. If the value of the partial derivative is close to 0, the final gradient to the transmitter will be very small. This gradient vanishing problem makes it difficult to train the transmitter through the multi-layer generator.} Inspired by the idea of residual learning~\cite{he1}, we intentionally construct a skip connection between the input and output layers of the generator, which is shown by the residual generator in Fig. \ref{FIG.3}. For the residual generator, the residual generating function $\bm{f}_{\bm{\theta_G^R}}: {\mathbb{C}}^n \mapsto {\mathbb{C}}^n$ could be denoted by
\begin{equation}\label{eq8}
\bm{f}_{\bm{\theta_G^R}}\left(\bf{x}\right)={\widetilde{\bf{y}}-\bf{x}}=\bm{f}_{\bm{\theta_G}}\left(\bf{x}\right)-\bf{x},
\end{equation}
where $\bf\bm{x}$ and $\bf\widetilde{y}$ are transmitted and generated signals, respectively, and $\bm{f}_{\bm{\theta_G^R}}\left(\bf{x}\right)$ is a residual generator aiming to learn the difference between transmitted and received signals with the conditional input $\bf{\bm x}$.
{\color{black}The distribution of the difference between transmitted and received signals is considered to be easier to learn than the distribution of the received signal ${p_{{h}}(\bf{y}|\bf{x})}$. For AWGN channel, in which the real received signal is denoted by ${\bf y}={\bf x}+{\bf n}$, the signal generated by residual generator $\bm{f}_{\bm{\theta_G^R}}$ is expected to close to Gaussian noise, while the traditional generator needs to approximate the whole received signal $\bf{y}$.} The gradient in each iteration for the transmitter DNN can be computed as:
\begin{align}\label{eq9}
\nabla_{\bm{\theta_T}}\widetilde{\mathcal{L}}\left(\bm{\theta_T}\right)
&=\frac{1}{B}\sum_{\mathit{i=}\mathrm{1}}^\emph{B}\frac{\partial l}{\partial \bm{f}_{\bm{\theta_R}}}\frac{\partial \bm{f}_{\bm{\theta_R}}}{\partial \bm{f}_{\bm{\theta_G}}}\frac{\partial \bm{f}_{\bm{\theta_G}}}{\partial \bm{f}_{\bm{\theta_T}}}\nabla_{\bm{\theta_T}}\bm{f}_{\bm{\theta_T}}\left({\bf{1}}_m^{\left(\mathit{i}\right)}\right)\notag \\
&=\frac{1}{B}\sum_{\mathit{i=}\mathrm{1}}^\emph{B}\frac{\partial l}{\partial \bm{f}_{\bm{\theta_R}}}\frac{\partial \bm{f}_{\bm{\theta_R}}}{\partial \bm{f}_{\bm{\theta_G}}}\frac{\partial \bm{f}_{\bm{\theta_G^R}}}{\partial \bm{f}_{\bm{\theta_T}}}\nabla_{\bm{\theta_T}}\bm{f}_{\bm{\theta_T}}\left({\bf{1}}_m^{\left(\mathit{i}\right)}\right) \notag \\
&\ \ \  +\frac{1}{B}\sum_{\mathit{i=}\mathrm{1}}^\emph{B}\frac{\partial l}{\partial \bm{f}_{\bm{\theta_R}}}\frac{\partial \bm{f}_{\bm{\theta_R}}}{\partial \bm{f}_{\bm{\theta_G}}}\nabla_{\bm{\theta_T}}\bm{f}_{\bm{\theta_T}}\left({\bf{1}}_m^{\left(\mathit{i}\right)}\right)\notag \\
&{\color{black}=\frac{1}{B}\sum_{\mathit{i=}\mathrm{1}}^{B}\frac{\partial l}{\partial \bf{p}^{(\mathit i)}}\frac{\partial \bf{p}^{(\mathit i)}}{\partial \bf{\widetilde y}^{(\mathit i)}}\frac{\partial{\bm{f}_{\bm{\theta_G^R}}}^{(\mathit i)}}{\partial \bf{x}^{(\mathit i)}}\nabla_{\bm{\theta_T}}\bm{f}_{\bm{\theta_T}}\left({\bf{1}}_m^{\left(\mathit{i}\right)}\right)} \notag\\
&\ \ \ {\color{black}+\frac{1}{B}\sum_{\mathit{i=}\mathrm{1}}^{B}\frac{\partial l}{\partial \bf{p}^{(\mathit i)}}\frac{\partial \bf{p}^{(\mathit i)}}{\partial \bf{\widetilde y}^{(\mathit i)}}\nabla_{\bm{\theta_T}}\bm{f}_{\bm{\theta_T}}\left({\bf{1}}_m^{\left(\mathit{i}\right)}\right),}
\end{align}
{\color{black}where the $\bm{f}_{\bm{\theta_R}}, \bm{f}_{\bm{\theta_G}}, \bm{f}_{\bm{\theta_G^R}}$, and $\bm{f}_{\bm{\theta_T}}$ are used to {\color{black}denote} the output of the receiver, generator, residual generator, and transmitter, respectively.} {\color{black}The training method of proposed RA-GAN is consistent with the traditional GAN \cite{ye1}. Specifically, we use the generated fake received signal ${\widetilde{\bf y}}$ and the real received signal ${{\bf y}}$ to train the discriminator of RA-GAN, and only used the generated fake received signal ${\widetilde{\bf y}}$ to train the generator of RA-GAN. Next, the receiver was only trained according to the real received signal ${{\bf y}}$, while the transmitter was only trained according to the fake received signal ${\widetilde{\bf y}}$. Among them, the loss function for training the transmitter and receiver are $l(\bm{f}_{\bm{\theta_R}}({\widetilde{\bf{y}}}),{\bf{1}}_m)$ and $l(\bm{f}_{\bm{\theta_R}}({\bf{y}}),{\bf{1}}_m)$, respectively.} Note that (\ref{eq9}) is the gradient for updating the transmitter DNN weights of the proposed RA-GAN based training scheme. There are two items on the right side of (\ref{eq9}). The first item is the same as that (\ref{eq6}), while the second item denotes the gradient through the skip connection between the input and output layers of the generator. Compared with the conventional GAN, the RA-GAN could generate a extra gradient to efficiently train the transmitter DNN due to the extra second item in (\ref{eq9}), thus the gradient vanishing problem could be mitigated. \footnote{\color{black}To obtain the optimal gradient, the derivative $\frac{\partial \bm{f}_{\bm{\theta_G^R}}}{\partial \bm{f}_{\bm{\theta_T}}}$ is expected to close the value of $h^{(i)}-1$, instead of ignoring this item. Thus, there is still a difference here when compared with the optimal case after adding the residual item. {\color{black}Specially, for the AWGN channel, i.e., $h^{(i)}=1$, since the derivative $\frac{\partial \bm{f}_{\bm{\theta_G^R}}}{\partial \bm{f}_{\bm{\theta_T}}}$ is expected to zero, the residual generator could be omitted.}}

It should be pointed out that for the RA-GAN based training scheme, only a small number of extra operations are needed both in feedforward calculation and backpropagation procedure due to the skip connection. Therefore, the extra computational complexity for the residual generator is negligible compared with the addition and multiplication operations for the conventional generator. 

{\color{black}Different from the existing residual-based generator used in image deblurring and image super-resolution~\cite{re_gen1,re_gen2,re_gen3,re_gen4,re_gen5}, which mainly focus on making full use of the low-level information by connecting with low layers to mitigate the gradient vanishing problem when training generator, the proposed residual generator is utilized to solve the gradient vanishing problem when training the transmitter. Moreover, compared to the complex design of the residual structure in~\cite{re_gen1,re_gen2,re_gen3,re_gen4,re_gen5}, the proposed RA-GAN is more straightforward by building a link back to the transmitter.

}

\IncMargin{1em} %
\begin{algorithm}
	\SetAlgoNoLine 
	
	\KwIn{
		\\
		1) Maximum number of iterations Epoch;\\
		2) Real channel dataset $\mathcal{H}$;\\}
	\KwOut{\\ \ \ \ Trained transmitter and receiver DNNs weights $\bm{\theta_T}$ and $\bm{\theta_R}$;\\
	}
	\BlankLine
	Initialization:  $ \ M=16, \ n=7, \ B=320, \ N_{\rm train}=10000,  $
	$\ \  \delta^2=(\frac{2E_b\log_2M}{N_0n})^{-1},\  \lambda=0.01,$ 
	$\ \ {\rm{Index}} = \lfloor N_{\rm train}/B\rfloor$\; 
	Generate $N_{\rm train}$ training data samples at random\;
	Initialize weights $\bm{\theta}_{\bm{D}},\bm{\theta}_{\bm{G}},\bm{\theta}_{\bm{R}}$, and $\bm{\theta}_{\bm{T}}$ according to the Xavier initialization method~\cite{Xavier1}\;
	\For{epoch = $1,2,\cdot\cdot\cdot,{\rm Epoch}$}{
		\For {index = $1,2,\cdot\cdot\cdot,{\rm Index}$}{
			Take $B$ one-hot vectors as training samples\;
			Caculate transmitted signals: ${\bf\bm{x}}^{\left(1\right)},\cdot\cdot\cdot,{\bf\bm{x}}^{(B)}$\;
			Take $B$ channel samples:
			${\bf\bm{h}}^{\left(1\right)},\cdot\cdot\cdot,{\bf\bm{h}}^{(B)}$\; 
			Get real received signals: ${\bf\bm{y}}^{\left(1\right)},\cdot\cdot\cdot,{\bf\bm{y}}^{(B)}$\;
			Generate fake received signals: ${\bf\bm{\widetilde y}}^{\left(1\right)},\cdot\cdot\cdot,{\bf\bm{\widetilde y}}^{(B)}$\;
			\For{$i\in\left\lbrace \bm{D,G,R,T} \right\rbrace $}{
				Caculate the loss function $\widehat{\mathcal{L}}\left(\bm{\theta}_{i}\right)$ according to (\ref{eq10})\;
				Use $\nabla_{\bm{\theta_{i}}}\widehat{\mathcal{L}}\left(\bm{\theta}_{i}\right)$ to update $\bm{\theta}_{i}$ according to the Adam method~\cite{adam}\;
			}
		}
	}
	Return $\bm{\theta}_{\bm{R}}$ and $\bm{\theta}_{\bm{T}}$.
	\caption{RA-GAN based E2E training scheme.\label{al1}}
\end{algorithm}
\DecMargin{1em}

\subsection{Regularization method to mitigate overfitting}
In this subsection, we reconstruct the loss function for the E2E learning of communication system to mitigate the overfitting problem. As the generator and the discriminator in GAN are added to train the E2E learning of communication system, the representation ability will substantially increase due to a mass of extra trainable DNN weights, which results in the overfitting problem~\cite{Good1}. {\color{black}To be specific, when a mass of parameters {\color{black}is} iteratively trained for the transmitter $\bm T$, receiver $\bm R$, generator $\bm G$, and discriminator $\bm D$, the residual generator is easy to overfit to partial training channel data. This overfitting problem results in residual generator performance degradation on other channel data.} To limit the representation ability of RA-GAN based training scheme, the regularizer is added in the loss function. Compared to the existing GAN based training scheme, the regularizer enables the RA-GAN based training scheme to generate a signal as similar to the real received signal as possible by using the regularization method.
Specifically, by adding a weight penalty item $\Omega\left(\bm{\theta}\right)$ in the original loss function in (\ref{eq5}) to restrict the representation ability of RA-GAN, we have 
\begin{equation}\label{eq10}
\widehat{\mathcal{L}}\left(\bm{\theta}_i\right) =\widetilde{\mathcal{L}}\bm{\left(\theta}_i\right)+ \lambda\Omega\left(\bm{\theta}_i\right), i\in\left\lbrace\bm{R,T,G,D} \right\rbrace,
\end{equation}
where $\widehat{\mathcal{L}}\left(\bm{\theta}_i\right)$ and $\widetilde{\mathcal{L}}\left(\bm{\theta}_i\right)$ are the reconstructed and original loss functions, respectively, $\lambda$ is the hyper-parameter to balance the penalty item and original loss function $\widetilde{\mathcal{L}}\left(\bm{\theta}_i\right)$.  ${\bm R}$, ${\bm T}$, ${\bm G}$, and ${\bm D}$ represent the receiver, transmitter, generator, and discriminator in the RA-GAN based training scheme, respectively. In this paper, we use $\bm l_2$ regularization $\frac{1}{2}||\bm\theta||^2$ as the penalty item. {\color{black}Note that the ${\bm l}_2$ regularizer could achieve better performance than the gradient penalty in WGAN loss and the ${\bm l}_1$ for weight sparsity, by avoiding large weights in the E2E system.} The key procedures of the RA-GAN based training scheme are described in $\bf {Algorithm}$ 1. {\color{black}Then, we {\color{black}aim} to minimize the reconstructed loss function (10), which makes each weight of $\bm\theta_i$ close to 0. Specifically, for a neural network (NN), if the weights are large, a small noise of the input data will have a great impact on the output results. But if the weights are small enough, it doesn't matter if the input data is shifted a little bit by noise. It is generally considered that the model with small weights values is relatively simple, and avoids overfitting problem~\cite{Good1}. Finally, we will obtain a well-trained residual generator without an overfitting problem, which could have a good performance in most channel data.}

{\color{black}As we all know, the mode collapse always happens in the GAN-based model in image processing and natural language processing. However, we have not observed the problems resulting from model collapse. The main reason is that the distribution of image and text is more complicated than the received signal distribution in the communication system. Thus, the mode of the received signal is easier to generate than image and text. Moreover, for the gradient vanishing problem when training generator. Specifically, if the discriminator could completely distinguish between real and fake data, i.e.,  $\bm{f}_{\bm{\theta_D}}({\bf \widetilde y})=0$ and $\nabla_{{\bf\widetilde y}}\bm{f}_{\bm{\theta_D}}({\bf \widetilde y})=0$, the gradients of the loss function to train the generator are close to zeros, where $\nabla_{{\bf\widetilde y}}\bm{f}_{\bm{\theta_D}}({\bf \widetilde y})=0$ denote the derivative of $\bm{f}_{\bm{\theta_D}}({\bf \widetilde y})$ is 0 in the neighborhood of ${\bf \widetilde y}$. On the one hand, we use the regularization method to make the discriminator $\bm D$ simple. On the other hand, we use the less learning rate to train the discriminator $\bm D$, and make it convergence slowly. In the end, the matching convergence rate between the discriminator $\bm D$ and generator $\bm G$ could be obtained to avoid the gradient vanishing problem when training the generator.}

Note that the penalty item $\Omega\left(\bm{\theta}_i\right)$ in (\ref{eq10}) only introduces limited operations when we compute the new loss function and gradient in each iteration. The computational complexity after reconstructing the loss function is $\mathcal{O}\left(|\bm{\theta_D}|+|\bm{\theta_G}|+|\bm{\theta_R}|+|\bm{\theta_T}|\right)$, where $|\bm{\theta}|$ denote the number of weights in $\bm{\theta}$. This increased complexity is still negligible compared with addition and multiplication operations in multi-layer DNN.

\section{Simulation Results}\label{S4}
In this section, we investigate the performance of the proposed RA-GAN based training scheme in terms of block error rate (BLER) for data transmission in the AWGN channel, Rayleigh channel, and DeepMIMO channel dataset based on ray-tracing~\cite{Alkhateeb2019}, respectively{\footnote{\color{black}It's worth noting that the proposed RA-GAN based training scheme is still suitable for non-linear channels, such as optical fiber channel~\cite{optical_fiber_channel}.}}. {\color{black}We compare the performance of the indirect RA-GAN based training scheme with indirect RL\cite{hoydis4}, indirect GAN based training scheme\cite{ye1}, indirect WGAN based training scheme\cite{wgan1}, and the direct optimal training method with known channel. In the direct optimal training method, we assume the real channel are known at the transmitter, which makes the gradient $\nabla_{\bm{\theta_T}}\widetilde{\mathcal{L}}$ available to train the transmitter DNN.} In addition, we analyze the ability of RA-GAN to generate a fake received signal and compare it with GAN. The layouts of the transmitter, receiver, generator, and discriminator are described in \textbf{Table \ref{Table1}}. Moreover, the parameter $E_b/N_0$ denotes the ratio of energy per bit ($E_b$) to the noise power spectral density ($N_0$), and the noise power $\delta^2$ equals to $(\frac{2E_b\log_2M}{N_0n})^{-1}$~\cite{hoydis1}. {\color{black}Expect the neural network dimension is shown in Table I, the training hyper-parameters are set as: weight decay $\lambda=0.01$, the maximum number of iteration Epoch $=200$, and the learning rate to train the transmitter and receiver is $0.001$. Specifically, for AWGN and Rayleigh fading channel model, the learning rate to {\color{black}train} the generator and discriminator are 0.0005 and 0.0001, respectively, and the batch size $B=320$. For the DeepMIMO channel model, due to the complex distribution of DeepMIMO, the learning rate to {\color{black}train} the generator and discriminator are set as 0.00005 and 0.00001, respectively, and the batch size $B=640$, to reduce the speed of the training model and make the model converge better.}

\renewcommand\arraystretch{1.2}
\begin{table}[!htbp]
	\footnotesize
	\color{black}
	\centering
	\caption{System parameters in RA-GAN based training scheme.}\label{Table1}
	\begin{tabular}{c|c|c|c} 
		\hline
		\    &Input  &Output  &Activation function \\  \hline
		\multirow{3}*{Transmitter }&$M$&2$M$&ReLU\\  
		\cline{2-4}  
		&2$M$&2$n$&Linear\\
		\cline{2-4}  
		&2$n$&2$n$&Normalization\\
		\cline{2-4} \hline
		\multirow{2}*{Receiver }&2$n$ / 4$n$&4$M$&ReLU\\  
		\cline{2-4}  
		&4$M$&$M$&Softmax\\
		\cline{2-4} \hline
		\multirow{3}*{Generator }&4$n$&8$M$&ELU\\  
		\cline{2-4}  
		&8$M$&8$M$&Tanh\\
		\cline{2-4}
		&8$M$&2$n$&Linear\\
		\cline{2-4} \hline
		\multirow{3}*{Discriminator }&2$n$ / 4$n$&2$M$&ELU\\  
		\cline{2-4}  
		&2$M$&2$M$&ELU\\
		\cline{2-4} 
		&2$M$&1&Sigmoid\\
		\cline{2-3} \hline
	\end{tabular}
\end{table}

\subsection{Generation capability comparison between RA-GAN and GAN}\label{4a}
At first, we compare the generation performance of the conventional GAN and the proposed RA-GAN for training the E2E communication system. Specifically, we consider the reconstructed loss function $\widehat{\mathcal{L}}\left(\bm{\theta_R}\right)$, $\widehat{\mathcal{L}}\left(\bm{\theta_T}\right)$ in (\ref{eq10}) for RA-GAN based training scheme and the original loss function $\widetilde{\mathcal{L}}\left(\bm{\theta_R}\right)$, $\widetilde{\mathcal{L}}\left(\bm{\theta_T}\right)$ in GAN based training scheme. As mentioned in Subsection \ref{2b}, the motivation of introducing RA-GAN and GAN is to generate a signal as similar to the real received signal as possible, i.e., the loss functions to train the transmitter DNN should be very close and consistent with the loss function to train the receiver DNN.
\begin{figure}[tp]
	\begin{center}
		\hspace*{0mm}\includegraphics[width=1\linewidth]{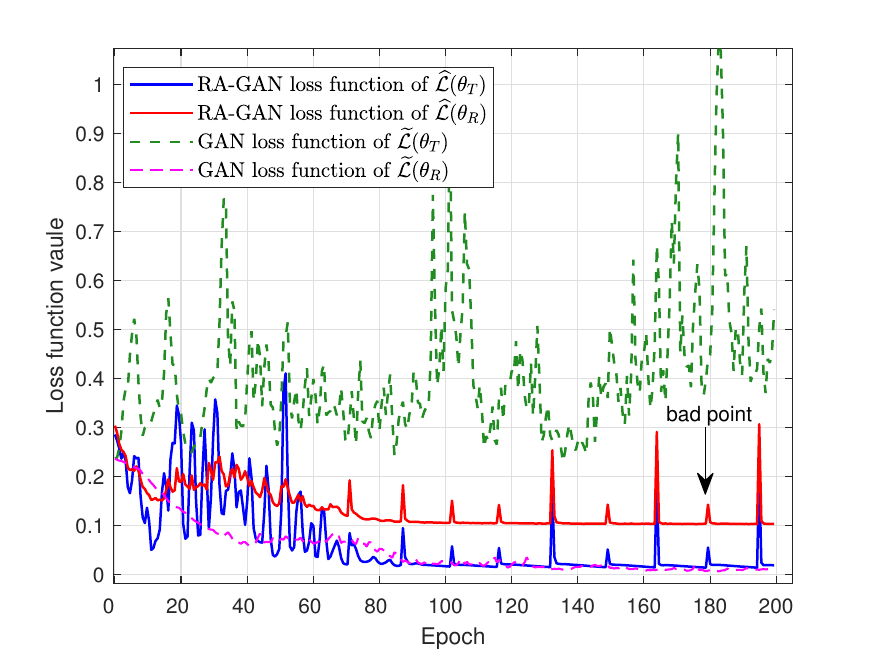}
	\end{center}
	\vspace*{-3mm}\caption{Generation performance comparison between the conventional GAN and the proposed RA-GAN.} \label{FIG.7}
	\vspace*{+2mm}
\end{figure}

In Fig. \ref{FIG.7}, we show the values of the corresponding loss functions against the training epoch in the AWGN channel, in which the received signal $\bf y$ can be expressed as $\bf y=\bf x+\bf w$, where $\bf x$ and $\bf{w}$ are the transmitted signal and Gaussian noise, respectively. The system is trained at $E_b/N_0=6$ dB. We can observe that the original loss function $\widetilde{\mathcal{L}}\left(\bm{\theta_T}\right)$ in the GAN based training scheme can not converge, and it is not close to the $\widetilde{\mathcal{L}}\left(\bm{\theta_R}\right)$. This result is caused by the gradient vanishing and overfitting problems. On the contrary, the reconstructed loss functions $\widehat{\mathcal{L}}\left(\bm{\theta_T}\right)$ and $\widehat{\mathcal{L}}\left(\bm{\theta_R}\right)$ in the proposed RA-GAN based training scheme are very stable, and they converge faster than the original loss functions $\widetilde{\mathcal{L}}\left(\bm{\theta_T}\right)$ and $\widetilde{\mathcal{L}}\left(\bm{\theta_R}\right)$. Note that in the training process, there are still some bad points due to the randomness of training, but the bad points could recover in the next epoch. The gap between $\widehat{\mathcal{L}}\left(\bm{\theta_R}\right)$ and $\widehat{\mathcal{L}}\left(\bm{\theta_T}\right)$ is equal to the gap between the penalty item $\lambda\Omega\left(\bm{\theta_R}\right)$ and $\lambda\Omega\left(\bm{\theta_T}\right)$. Thus, compared with the existing GAN based training scheme, the proposed RA-GAN based training scheme could generate a much more similar signal to the real received signal, which shows that the trained residual generator has better generation performance than the conventional generator.
\subsection{BLER performance in the AWGN channel}\label{4b}

Next, we compare the performance of the RA-GAN based training scheme, GAN based training scheme\cite{ye1}, the RL based training scheme\cite{hoydis4}, and the optimal training method in the AWGN channel{\footnote{{\color{black}There is no channel estimation module~\cite{dai2} to estimate the channel coefficient in E2E communication system. So, we verify the performance of the proposed RA-GAN based training scheme and other training schemes in the AWGN channel model and other channel models without known channel.}}}. The training parameters are the same as those in Subsection \ref{4a}. As mentioned in Section \ref{S3a}, the residual generator just needs to generate Gaussian output. This is simple to realize by scaling the standard Gaussian input $\bf z$. 

\begin{figure}[tp]
	\begin{center}
		\hspace*{0mm}\includegraphics[width=1\linewidth]{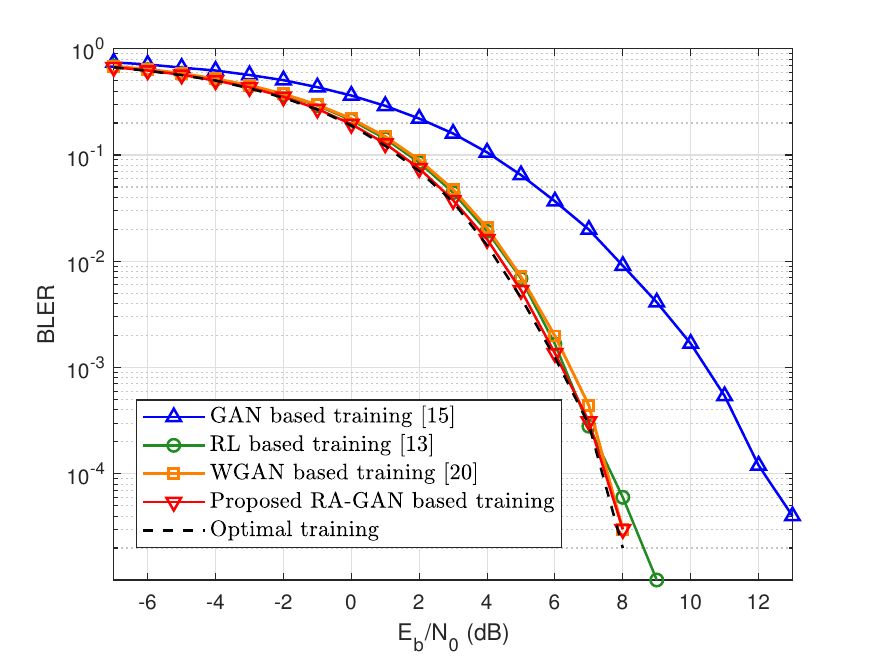}
	\end{center}
	\vspace*{-4mm}\caption{BLER performance comparison in the AWGN channel.} \label{FIG.4}
	\vspace*{+1mm}
\end{figure}

{\color{black}To compare the performance of different training schemes in larger $E_b/N_0$ scope as done in~\cite{hoydis1}, we test the BLER performance with a validation dataset including 100,000 random one-hot vectors from -7 dB to 13 dB.} As shown in Fig. \ref{FIG.4}, we can observe that the BLER performance gap between the existing GAN based training scheme~\cite{ye1} and the optimal training scheme with known channel is large. This large performance gap is caused by the gradient vanishing and overfitting problems when training GAN. On the contrary, the proposed RA-GAN based training scheme almost approaches the optimal training method.

\subsection{BLER performance in the Rayleigh fading channel}\label{4c}
In this subsection, we consider the Rayleigh fading channel ${{h}}\sim\mathcal{CN}\left({0,1}\right)$. The receiver signal $\bf{y}$ could be denoted by ${\bf{y}}={h}\bf{x}+w$. {\color{black}Unlike the AWGN channel, some known pilot signals ${\bf x_p}$ are transmitted to help to train the E2E learning of communication system in the Rayleigh fading channel and DeepMIMO channel. The received pilot signal ${\bf y_p}$ concatenate with received data signal ${\bf y}$ as the input of receiver and discriminator of RA-GAN, GAN, and WGAN.}

\begin{figure}[tp]
	\begin{center}
		\hspace*{0mm}\includegraphics[width=1\linewidth]{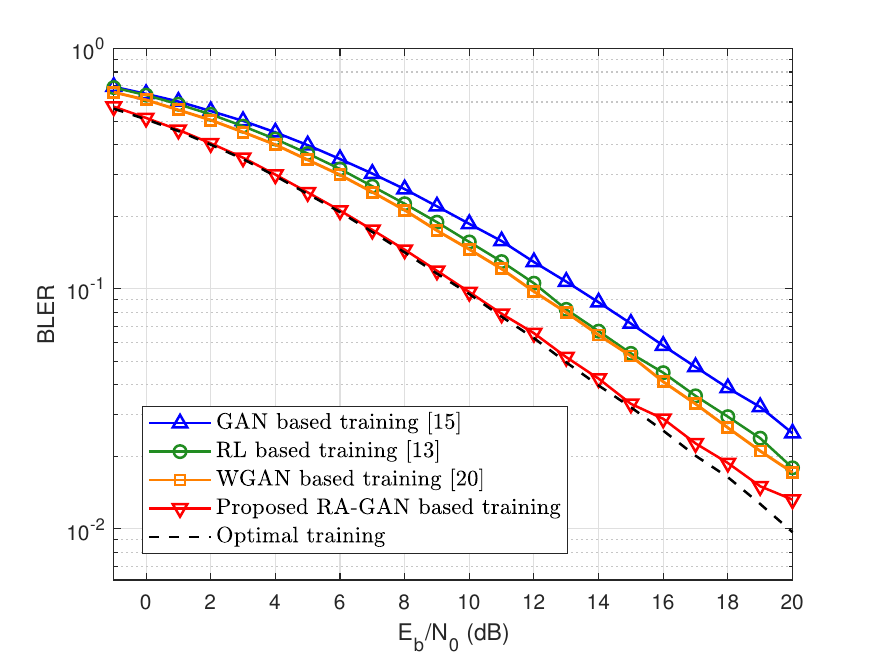}
	\end{center}
	\vspace*{-4mm}\caption{BLER performance comparison in the Rayleigh fading channel.} \label{FIG.5}
	\vspace*{+1mm}
\end{figure}

In Fig. \ref{FIG.5}, we show the BLER performance of RA-GAN based training scheme compared with GAN based training schemes\cite{ye1} and RL based training schemes\cite{hoydis4} in Rayleigh fading channel. We train the system at $E_b/N_0=16$ dB, and double the input dimension of the receiver and RA-GAN due to the use of the pilot. {\color{black}We observe that the RA-GAN based training scheme outperforms the RL, WGAN, and GAN based training schemes, e.g., the proposed RA-GAN based training scheme outperforms the GAN based training scheme by 3 dB when the BLER is 0.1. Since the WGAN is designed to solve the mode collapse problem which doesn’t appear in E2E training, compared to the GAN method, WGAN based training scheme only achieves smaller performance improvement.} When $E_b/N_0$ is lower than 15 dB, the BLER performance of RA-GAN based training scheme could almost overlap with the optimal training method. When $E_b/N_0$ is greater than 16 dB, the BLER performance of the proposed RA-GAN based training scheme cannot approach the optimal training method. This is caused by the fact that the regularizer will limit the representation ability of the E2E learning of communication system. If we decrease the hyper-parameter $\lambda$ in (\ref{eq10}), the optimal BLER performance will be achieved, but the BLER performance will degrade in low $E_b/N_0$ regions.

\subsection{BLER performance in the ray-tracing based DeepMIMO dataset}\label{4d}

To verify the performance of the proposed RA-GAN based training scheme in the real channel, we use the DeepMIMO channel dataset based on ray-tracing~\cite{Alkhateeb2019} to generate channel samples which is more realistic than the simulated channel model, e.g., the AWGN channel and the Rayleigh fading channel. By using Wireless InSite ray-tracing simulator~\cite{remcom1}, the DeepMIMO channel dataset can capture the dependence on the real channel environmental factors such as user location and environment geometry and so on. One main advantage of the DeepMIMO channel dataset is that the dataset could be completely defined by the parameters set and the 3D ray-tracing scenario. In our simulations, the DeepMIMO channel data are generated according to the parameters shown in \textbf{Table \ref{Table2}}. This DeepMIMO channel dataset contains the channels between the BS 3 deployed with single-antenna antenna and single-antenna users from Row 1 to Row 2751, which is divided into a training set and a validation set. The training set with 80\% of the data is used to train the E2E learning of communication system, while the validation set with the rest 20\% data is used to test the system performance.


\renewcommand\arraystretch{1.2}
\begin{table}
	\centering
	\caption{The DeepMIMO dataset simulation parameters.}\label{Table2}
	\begin{tabular}{|l|l|}
		\hline
		
		Parameter & Value \\ \hline
		Name of scenario & O1 \\ \hline
		Active BS & 3 \\ \hline
		Active users & Row 1-2751  \\ \hline
		Number of users & 497,931 \\ \hline
		Number of BS antenna in (x,y,z) & (1,1,1) \\ \hline
		System bandwidth & 0.5 GHz \\ \hline
		System spectrum & 60 GHz \\ \hline
		Number of OFDM sub-carriers & 64 \\ \hline
		OFDM limit & 1 \\ \hline
		Number of channel paths & 5 \\ \hline
		
	\end{tabular}
\end{table}

\begin{figure}[tp]
	\begin{center}
		\hspace*{0mm}\includegraphics[width=1\linewidth]{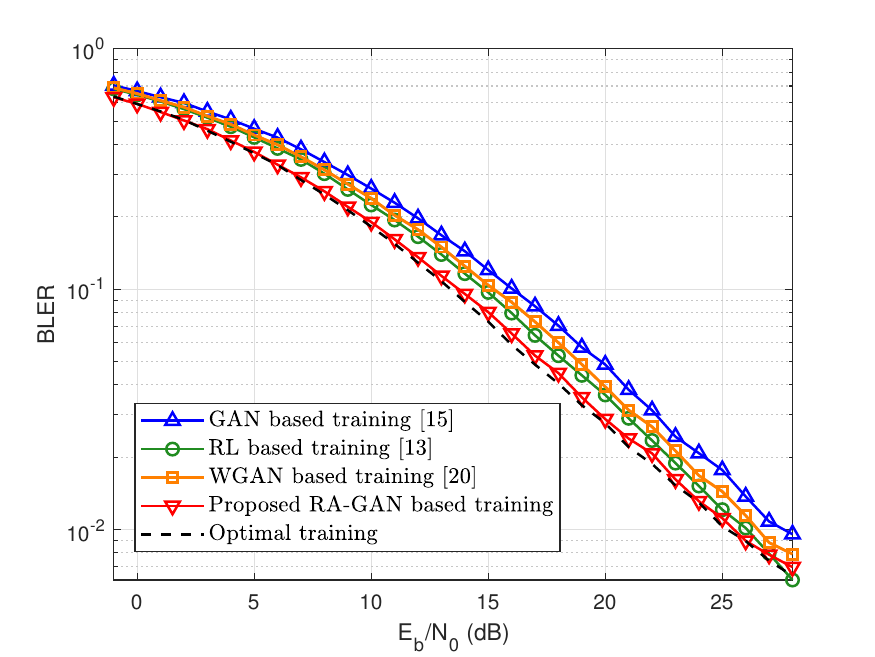}
	\end{center}
	\vspace*{-4mm}\caption{BLER performance comparison in DeepMIMO channel dataset.} \label{FIG.6}
	\vspace*{+1mm}
\end{figure}

 In Fig. \ref{FIG.6}, we show the BLER performance of RA-GAN based training scheme compared with GAN based training schemes\cite{ye1} and RL based training schemes\cite{hoydis4} in the DeepMIMO channel dataset. The training $E_b/N_0$ is 16 dB, the batch size is 640, and the hyper-parameter $\lambda$ is 0.005. We can observe that the proposed RA-GAN based training scheme outperforms the GAN and RL based training schemes, e.g., the proposed RA-GAN based training scheme outperforms the GAN based training scheme by 2 dB when the BLER is 0.1. Moreover, the proposed RA-GAN based training scheme can mitigate the gradient vanishing and overfitting problems of GAN, and thus achieve the near-optimal BLER performance.

{\color{black}
\subsection{Training time and complexity analysis}\label{4e}
We compare the training time of the methods for end-to-end training in different channel {\color{black}models}, as shown in \textbf{Table \ref{Table01}}. From this table, we can observe that the proposed RA-GAN based training scheme requires only a little more time than the GAN based training scheme. Although the RL based training scheme has a shorter training time, it requires the transmitter to transmit a large amount of extra signals for training. Moreover, we also calculate the complexity of the GAN, RA-GAN, and WGAN methods by comparing the number of weights in generator and discriminator, as shown in \textbf{Table \ref{Table02}}. Since the residual connection does not increase the number of weights, we can observe that the proposed RA-GAN based training scheme and GAN based training scheme have {\color{black}the} same amount of weights, both of which realize lower complexity compared with WGAN based training scheme.}

\renewcommand\arraystretch{1.2}
\begin{table}[!htbp]
	\footnotesize
	\color{black}
	\centering
	\caption{Training time comparison (sec).}\label{Table01}
	\begin{tabular}{c|c|c|c} 
		\hline
		\    &\ \ \ \ AWGN\ \ \ \  &Rayleigh fading &DeepMIMO \\  \hline
		\multirow{1}*{optimal}&35.12&27.15&16.28\\  
		\cline{2-4}  
		\hline
		\multirow{1}*{GAN~\cite{ye1}}&222.01&113.23&71.13\\  
		\cline{2-4}  
		\hline
		\multirow{1}*{WGAN~\cite{wgan1}}&1055.58&309.46&244.58\\  
		\cline{2-4} \hline
		\multirow{1}*{RA-GAN}&226.91&118.70&73.71\\  
		\cline{2-4} \hline
		\multirow{1}*{RL~\cite{hoydis4}}&61.08&60.82&37.14\\  
		\cline{2-4} \hline
	\end{tabular}
\end{table}

\renewcommand\arraystretch{1.2}
\begin{table}[!htbp]
	\footnotesize
	\color{black}
	\centering
	\caption{The number of weights in generator and discriminator, respectively.}\label{Table02}
	\begin{tabular}{c|c|c|c} 
		\hline
		\    &\ \ \ \ AWGN\ \ \ \  &Rayleigh fading &DeepMIMO \\  \hline
		\multirow{1}*{GAN~\cite{ye1}}&22030\ /\ 513&23822\ /\ 2017&23822\ /\ 2017\\  
		\cline{2-4}  
		\hline
		\multirow{1}*{WGAN~\cite{wgan1}}&26208\ /\ 13176&28096\ /\ 15512&28096\ /\ 15512\\  
		\cline{2-4} \hline
		\multirow{1}*{RA-GAN}&22030\ /\ 513&23822\ /\ 2017&23822\ /\ 2017\\  
		\cline{2-4} \hline
	\end{tabular}
\end{table}

{\color{black}
\subsection{BLER performance in the non-linear channel model}\label{4f}
Furthermore, we verify the performance of the proposed RA-GAN based training scheme in a non-linear channel model, i.e., optical fiber channel model. In an optical fiber communication system, the Mach-Zehnder modulator (MZM) is used to modulate electrical signals into optical signals in the transmitter, while simple photodiodes (PDs) are used to detect the intensity of the received optical field and perform the opto-electrical conversion in the receiver~\cite{optical_fiber_channel}. Specifically, the modulation process in the MZM modulator is modeled by
$m(t)=1+e^{jx(t)}$, where $x(t)$ is the $t$-th transmitted signal and $m(t)$ is the $t$-th modulated signal. Then, the modulated signal is transmitted through optical fiber, where fiber dispersion affects the signal quality. The fiber dispersion can be solved analytically in the frequency domain by taking the Fourier transform. Thus, the FFT and IFFT are necessary for conversion between the time and frequency domain, which can be realized by the Torch library. Finally, the received intensity of the optical field can be denoted by $r(t)=|h\{m(t)\}|^2+n(t)$, where $h\{\cdot\}$ is an operator describing the effects of the fiber dispersion and $n(t)$ is the additive white Gaussian noise. As a consequence of the joint effects of dispersion and intensity of the received optical field detection, the optical fiber communication channel is nonlinear.

\begin{figure}[t]
	\begin{center}
		\includegraphics[width=1\linewidth]{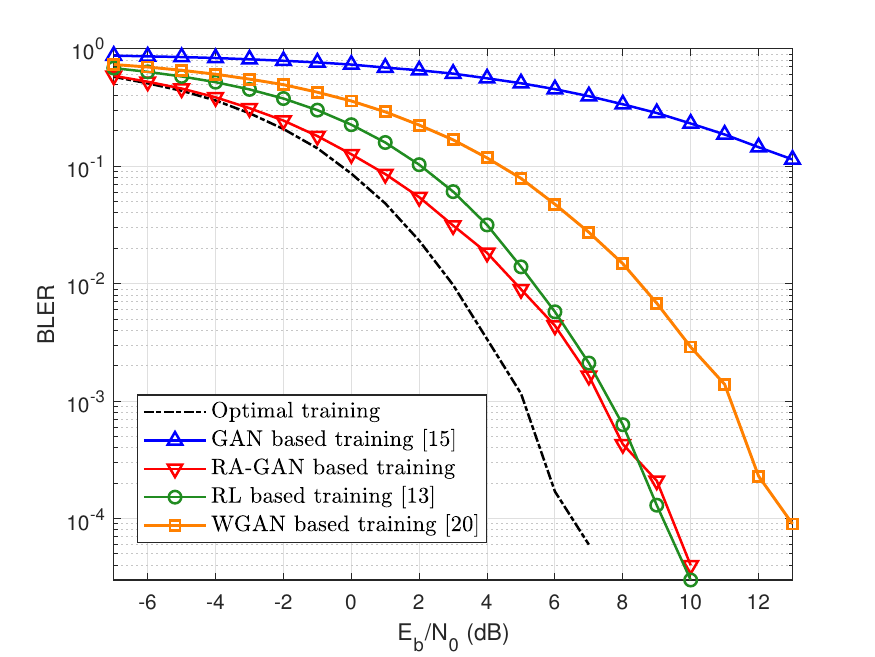}
	\end{center}
	\caption{BLER performance comparison in optical fiber channel.}\label{Fig8}
\end{figure}

 According to the optical fiber channel model introduced above, we compare the performance of the RA-GAN based training scheme, GAN-based training scheme, WGAN based training scheme, the RL-based training scheme, and the optimal training method in the optical fiber channel. We test the BLER performance with a validation dataset including 100,000 random one-hot vectors. As shown in Fig. \ref{Fig8}, we can observe that the proposed RA-GAN based training scheme outperforms the GAN and WGAN based training scheme, which illustrates the proposed RA-GAN has a better ability to characterize nonlinear channels. At the same time, RA-GAN based method achieves the best BLER performance in the low $E_b/N_0$ area, while RA-GAN based method is still competitive in the high $E_b/N_0$ area.
}

\section{Conclusions}\label{S5}
In this paper, we proposed the RA-GAN based training scheme for the E2E learning of communication system to train transmitter without a known channel. Specifically, we improved the surrogate gradient method by using residual learning to transform the conventional GAN into RA-GAN with a negligible increase in computational complexity. Based on the proposed RA-GAN based training scheme, more powerful and robust gradients can be achieved to solve the gradient vanishing problem. Furthermore, a regularizer was utilized in the RA-GAN to limit the representation ability, which can solve the overfitting problem. Simulation results verified the near-optimal BLER performance of the proposed RA-GAN based training scheme, which outperforms other deep learning methods in the AWGN channel, Rayleigh fading channel, and DeepMIMO channel dataset. For future research of E2E learning of communication system, we will focus on how to train the transmitter without a known channel in multiple-input multiple-output (MIMO) and multi-user scenarios.
\bibliography{IEEEabrv,Hao1Ref}

\begin{IEEEbiography}[{\includegraphics[width=1in,height=1.25in,clip,keepaspectratio]{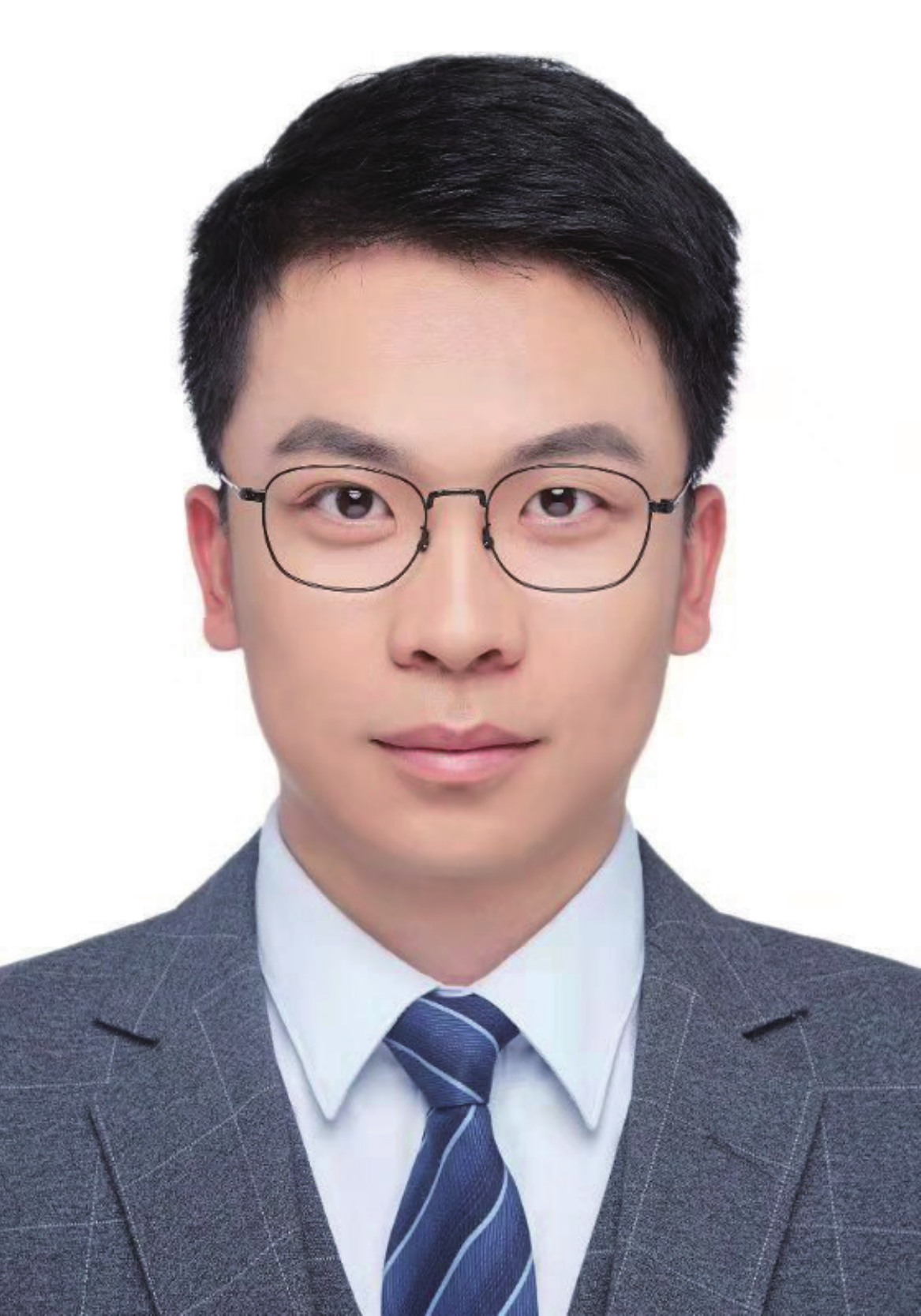}}]{Hao Jiang} (Student Member, IEEE) received the B.S. degree in physics from Tsinghua University, Beijing, China, in 2018, where he is currently pursuing the Ph.D. degree with the Department of Electronic Engineering. His research interests include	mmWave communications, machine learning for wireless communications, and reconfigurable intelligent surface (RIS).
\end{IEEEbiography}

\begin{IEEEbiography}[{\includegraphics[width=1in,height=1.25in,clip,keepaspectratio]{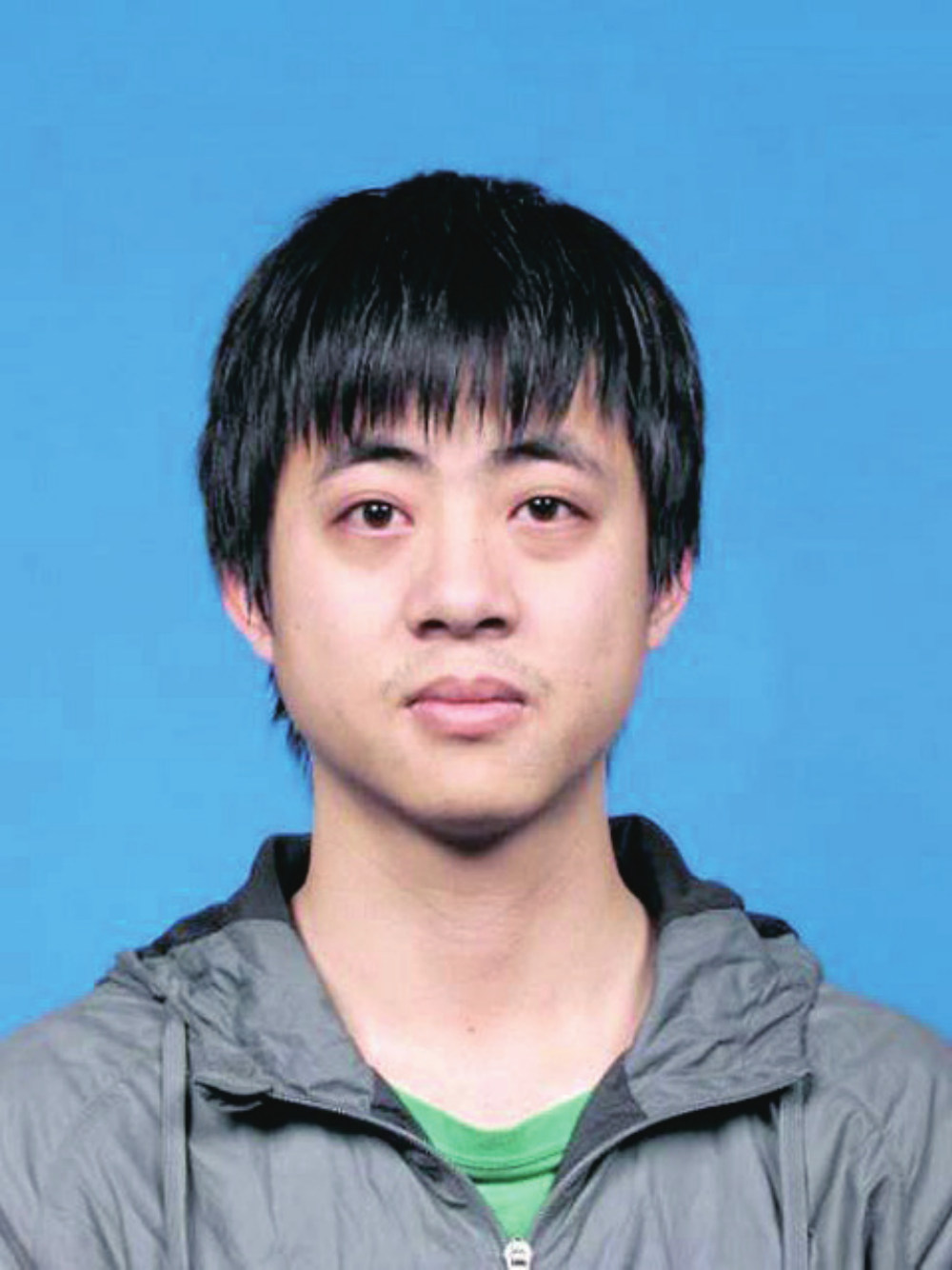}}]{Shuangkaisheng Bi} received the B.S. and M.S. degree from the Department of Electronic Engineering, Tsinghua University, in 2018 and 2021. Now, he is a Research Scientist at Huawei Technology. His research interests include array signal processing and radar signal processing. Besides, he is also interested in prototype development for advanced wireless communications.
\end{IEEEbiography}

\begin{IEEEbiography}[{\includegraphics[width=1in,height=1.25in,clip,keepaspectratio]{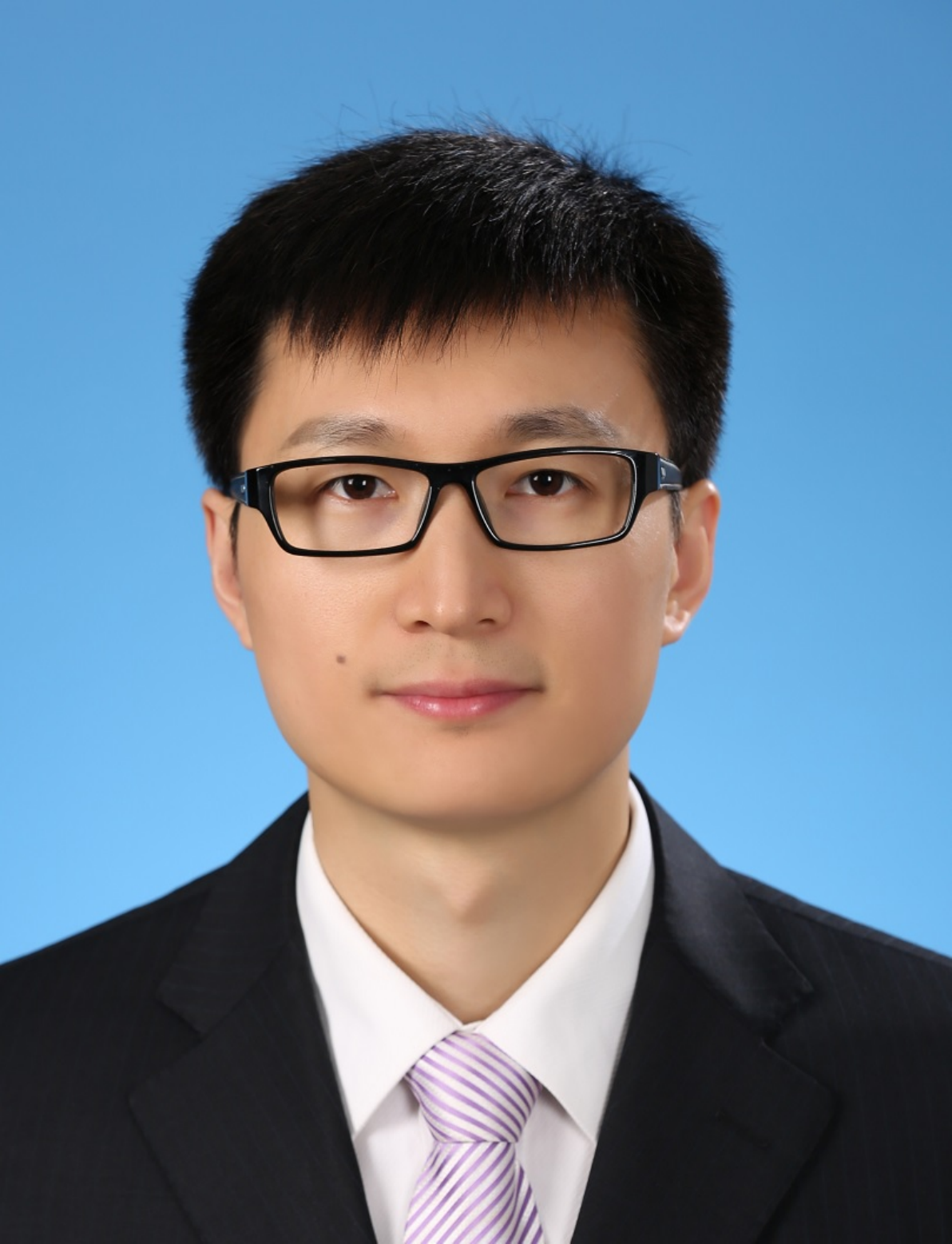}}]{Linglong Dai} (Fellow, IEEE) received the B.S. degree from Zhejiang University, Hangzhou, China, in 2003, the M.S. degree (with the highest honor) from the China Academy of Telecommunications Technology, Beijing, China, in 2006, and the Ph.D. degree (with the highest honor) from Tsinghua University, Beijing, China, in 2011. From 2011 to 2013, he was a Postdoctoral Research Fellow with the Department of Electronic Engineering, Tsinghua University, where he was an Assistant Professor from 2013 to 2016 and has been an Associate Professor since 2016. His current research interests include reconfigurable intelligent surface (RIS), massive MIMO, millimeter-wave and Terahertz communications, and machine learning for wireless communications.
	
	He has coauthored the book {\it MmWave Massive MIMO: A Paradigm for 5G} (Academic Press, 2016). He has authored or coauthored over 70 IEEE journal papers and over 40 IEEE conference papers. He also holds 19 granted patents. He has received five IEEE Best Paper Awards at the IEEE ICC 2013, the IEEE ICC 2014, the IEEE ICC 2017, the IEEE VTC 2017-Fall, and the IEEE ICC 2018. He has also received the Tsinghua University Outstanding Ph.D. Graduate Award in 2011, the Beijing Excellent Doctoral Dissertation Award in 2012, the China National Excellent Doctoral Dissertation Nomination Award in 2013, the URSI Young Scientist Award in 2014, the IEEE Transactions on Broadcasting Best Paper Award in 2015, the Electronics Letters Best Paper Award in 2016, the National Natural Science Foundation of China for Outstanding Young Scholars in 2017, the IEEE ComSoc Asia-Pacific Outstanding Young Researcher Award in 2017, the IEEE ComSoc Asia-Pacific Outstanding Paper Award in 2018, the China Communications Best Paper Award in 2019, the IEEE Access Best Multimedia Award in 2020, and the IEEE Communications Society Leonard G. Abraham Prize in 2020. He was listed as a Highly Cited Researcher by Clarivate Analytics in 2020 and 2021. He was elevated as an IEEE Fellow in 2022.
	
	He is an Area Editor of {\sc IEEE Communications Letters}, and an Editor of {\sc IEEE Transactions on Communications} and {\sc IEEE Transactions on Vehicular Technology}. Particularly, he is dedicated to reproducible research and has made a large amount of simulation codes publicly available.
	
\end{IEEEbiography}

\begin{IEEEbiography}[{\includegraphics[width=1in,height=1.25in,clip,keepaspectratio]{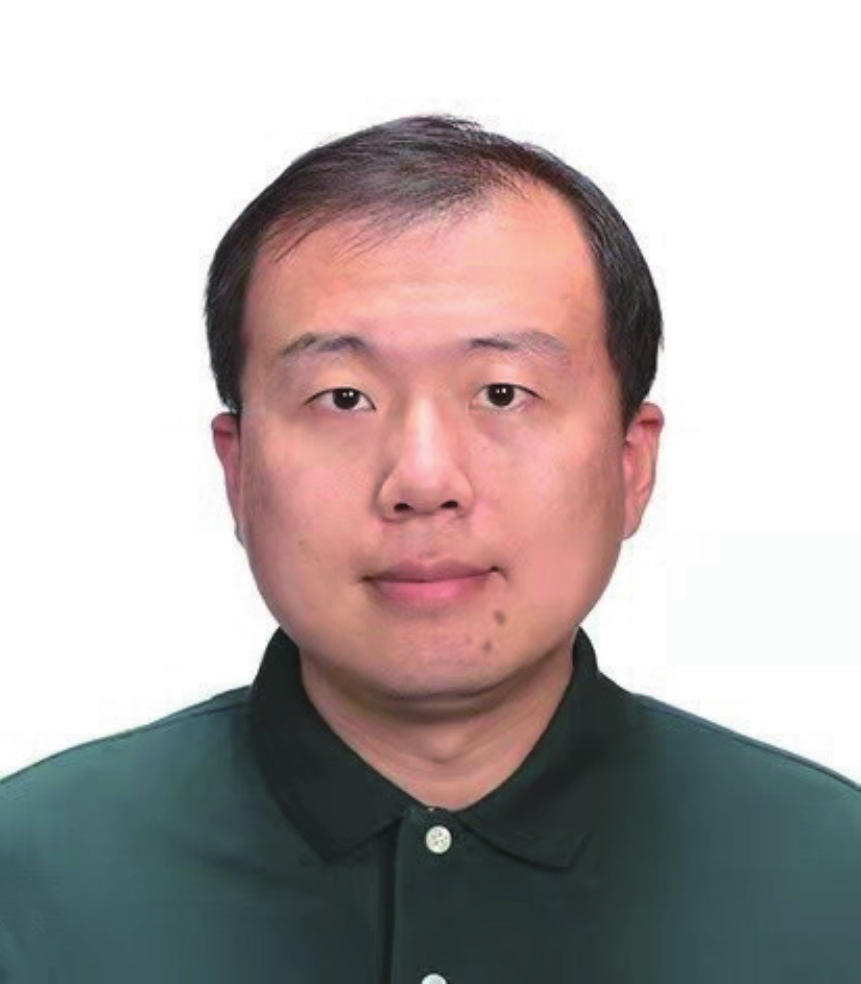}}]{Hao Wang} (Member, IEEE) received the Ph.D. degree in information and communication engineering from Tsinghua University, China, in 2010, and the Ph.D. degree in electronic engineering from the City University of Hong Kong, Hong Kong SAR, in 2011. He joined Huawei Technologies Company, Ltd., Beijing, in 2010, and is currently leading algorithm research and development for the Smart and Easy Receiver.
\end{IEEEbiography}

\begin{IEEEbiography}[{\includegraphics[width=1in,height=1.25in,clip,keepaspectratio]{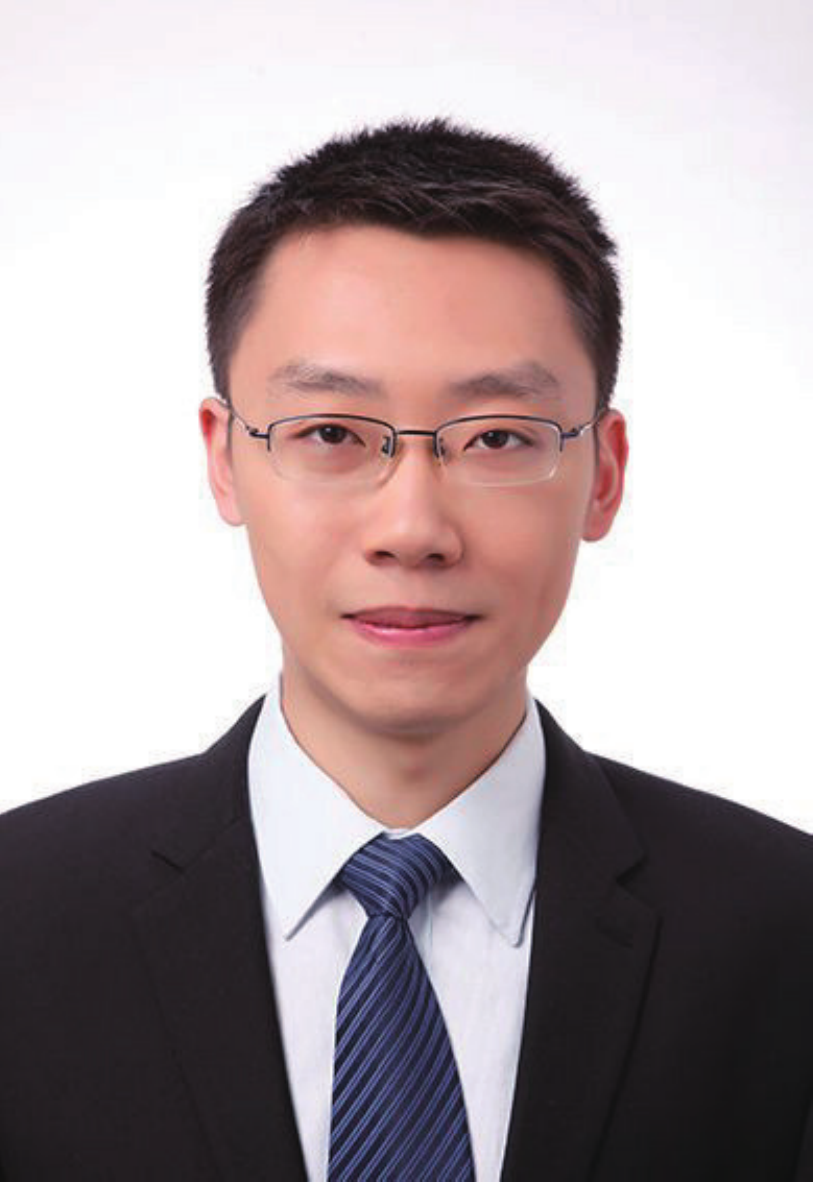}}]{Jiankun Zhang} (Member, IEEE) received the B.E. degree from the Beijing Institute of Technology in Beijing, China, in 2013, and Ph.D. degree in the signal and information processing from Peking University, Beijing, China, in 2018. From 2019, he serves as a researcher in Huawei Technologies Co., Ltd. His research interests include signal processing in 5G and 6G systems, AI based smart communications, mmWave communications, and wireless optical communications.
\end{IEEEbiography}

\end{document}